\documentclass[showpacs,amsmath,amssymb,amsfonts,a4paper,aps,prd]{revtex4}
\usepackage[colorlinks, linkcolor={RoyalBlue},citecolor={Red}]{hyperref}
\usepackage{graphicx,color}
\usepackage{float}
\usepackage[utf8]{inputenc}
\usepackage{multirow}
\usepackage[dvipsnames]{xcolor}

\def\be{\begin{equation}}
\def\ee{\end{equation}}
\def\bea{\begin{eqnarray}}
\def\eea{\end{eqnarray}}
\def\bl{\begin{align}}
\def\el{\end{align}}

\def\e{\mathrm{e}}

\begin{document}

\title{Warm-tachyon Gauss-Bonnet inflation in the light of Planck 2015 data }

\author{Meysam Motaharfar}
\email{mmotaharfar2000@gmail.com}
\author{Hamid Reza Sepangi$^1$}
\email{hr-sepangi@sbu.ac.ir}
\affiliation{$^1$Department of Physics, Shahid Beheshti University, G. C., Evin,Tehran 19839, Iran}

\begin{abstract}

 We study a warm-tachyon inflationary model non-minimally coupled to a Gauss-Bonnet term. The general conditions required for reliability of the model are obtained by considerations of a combined hierarchy of Hubble and Gauss-Bonnet flow functions. The perturbed equations are comprehensively derived in the longitudinal gauge in the presence of slow-roll and quasi-stable conditions. General expressions for  observable quantities of interest such as the tensor-to-scalar ratio, scalar spectral index and its running are  found in the high dissipation regime. Finally, the model is solved using  exponential and inverse power-law potentials, which satisfy the properties of a tachyon potential, with parameters of the model being constrained within the framework of the Planck 2015 data. We show that the Gauss-Bonnet coupling constant controls termination of inflation in such a way as to be in good agreement with the Planck 2015 data.

\end{abstract}

\date{\today}

\maketitle

                                                                          \section{Introduction}\label{sec1}

Observational data in the past few decades have brought about an elegant paradigm to describe  dynamics of the early universe, known today as inflation which naturally accounts for a number of long-standing problems of the standard Big Bang model including flatness, horizon and relic, to name but a few \cite{g1, g9}. However, a noteworthy feature of inflation is that it generates a mechanism to seed the Large-Scale Structure (LSS) of the universe \cite{q1} and also provides a causal interpretation for the origin of temperature anisotropies seen in the Cosmic Microwave Background (CMB) radiation \cite{q6}, traceable  to primordial density perturbations which may have been produced from quantum fluctuations during the inflationary era.

As is well known, during the inflationary era, the inflaton field undergoes a slow-roll period which is necessary for inflation to happen.
Broadly speaking, we might consider two main competing scenarios for slow-roll inflation; the first is the conventional supercooled inflation (isentropic) and the second is warm inflation (non-isentropic). During the standard inflation the universe undergoes two stages, a first order phase transition during which its temperature  decreases abruptly and therefore the inflaton field is assumed to be isolated and the interaction between the inflaton and other fields are neglected and, the second stage where due to this supercooling phase the universe enters a reheating epoch to get hot again and filled with  radiation required by the Big Bang scenario. The general consensus today is that the primordial quantum fluctuations are responsible to seed LSS in such models.

Warm inflation, as a complementary version of  standard inflation, was first proposed in \cite{q9} by Berera and Fang in which meshing these two isolated stages would resolve the disparities created by each separately. In the warm inflationary scenario, the accelerating universe is still driven by the potential energy density as in standard inflation, but because of interaction between the inflaton field and other fields, the radiation cannot be red-shifted strongly and the universe remains hot during inflation. During this period, the dissipative effects are significant so that radiation  occurs concurrently  with the inflationary expansion. The dissipating effects arise from a friction term which
describes the processes of the scalar field dissipating to a thermal bath. In fact, the radiation dominates immediately as soon as inflation ends. Since, thermal fluctuations are responsible to seed LSS instead of quantum fluctuations, this warm scenario will bring novel properties at late times. In addition, the matter ingredients of the universe are produced by the decay of either the remaining inflaton field or the dominant radiation field \cite{g16}. 
As a result, this scenario not only solves problems which the conventional inflationary scenario does, but  has additional advantages as follows: I-  thermal fluctuations during inflation may play a dominant role in producing the initial fluctuations which are indispensable for the LSS formation \cite{q9, g14,g15}, II- the slow-roll conditions can be satisfied even for steeper scalar potentials, III- the inflationary phase smoothly terminates and enters  a radiation dominated era and, in fact, the slow-roll and reheating periods are unified, IV- it may contribute a very interesting mechanism for baryogenesis, where the spontaneous baryo/leptogenesis can easily be realized in this scenario \cite{g27}, V- in regimes relevant to observation, the mass of inflaton is typically much larger than the Hubble scale and therefore this scenario does not suffer from the so-called eta problem \cite{q10}, VI- since the macroscopic dynamics of the background field and fluctuations are classical from the onset, there is no quantum-classical transition problem and finally, accounting for dissipative effects may be important in alleviating the initial condition problem of inflation \cite{q11}. 

In the recent past, it has been shown that tachyon fields associated with unstable D-branes may be responsible for inflation at early times \cite{g29} and can be an appropriate candidate for dark matter evolution during intermediate epoch \cite{q12}. As Gibbons has shown \cite{Gibbons}, if a tachyon condensate starts to roll down the potential with small initial velocity, then a universe dominated with new form of matter will smoothly evolve from an accelerated phase to an era dominated by a non-relativistic fluid, which could contribute to the dark matter mentioned above. Generally speaking, the tachyon field potentials have a maximum at $\phi = 0$ and a minimum at $\phi \rightarrow \infty$. There are then two types of potential satisfying these two conditions; an exponential potential ($V(\phi) = V_{0} e^{- \alpha \phi}$) and an inverse power law potential ($V(\phi) = V_{0} \phi^{-n}$). Due to such interesting characteristics, many studies have been made exploring tachyon inflationary models \cite{k1}.

In any study concerning the inflationary universe, quantum gravitational effects ought to be taken into consideration.  It is also believed that in the low-energy limit, string theory,  of which quantum gravity is a consequence,  corresponds to General Relativity including quadratic terms in the action. Furthermore, to have a ghost-free action, the Einstein-Hilbert action should have quadratic curvature corrections which would be proportional to a Gauss-Bonnet term which has topologically no contribution in 4 dimensions, except when coupled to other fields including scalar fields. In addition, this term has no problem with unitarity and since the equations of motion do not contain higher than second order in temporal derivatives, there would be no stability problem \cite{g30,g31,g32,g33,g34}. Fortunately, the theory with a non-minimally coupled Gauss-Bonnet term could provide the possibility of avoiding the initial singularity of the universe \cite{nn1}. It may violate the energy conditions thanks to the presence of a term in the singularity theorem \cite{nn2}. Therefore, such a quadratic term plays a significant role in the early universe dynamics.  In this prospective, it would  be of interest to study models where the Gauss-Bonnet term is directly coupled to a scalar field and study its effects in the early universe \cite{h1}.

To explore the viability of an inflationary model, properties of the initial cosmological perturbations play a vital role. Such properties will mainly be described with statistical parameters like the two point correlation function known as the  power spectrum, scalar and tensor spectral index, their running and tensor-to-scalar ratio. Having such parameter for a particular inflationary model gives the opportunity to check its viability using observational constraints. In this respect, several collaborations have tried to obtain new observational constraints on space parameters using recently released Planck data \cite{n1}. As a matter of fact, joining Planck likelihood with TT, TE and EE polarization modes give $n_{s} = 0.96435 \pm 0.00955$, $\alpha_{s} = -.00885\pm 0.01505$ and $r < 0.1488$ and adding BAO likelihood data constrain the space parameters to $n_{s} = 0.9656 \pm 0.00825$, $\alpha_{s} = -.00885\pm 0.01505$ and $r < 0.1504$ at $ 95\%$ confidence level.

Tachyon warm inflationary models and their perturbations have been studied in \cite{g35,g39}. Having the above points in mind, we build on the work of Herrera,  Del campo and Campuzano on warm-tachyon inflation \cite{g39} by adding a Gauss-Bonnet correction. We organize the paper by presenting our model in the next section and derive the flow functions \cite{g43,g46} and the number of e-folding,  followed by studying perturbations of this model in section III. In section IV,  we calculate the power spectrum generally and derive modified spectral index and tensor-to-scalar ratio of the model in the high dissipation regime in section V. In section VI and VII, we consider the above two mentioned potentials and analytically solve the model and obtain observable quantities in terms of the e-folding. In section VIII we consider further general functions and numerically solve the model. Through sections VI to VIII, we attempt to constrain our theoretical predictions by the Planck data, compare our results with the case where the Gauss-Bonnet term has no contribution and illuminate the characteristics of the model investigating the impact of the free parameters in a qualitative manner.

                                                                                 \section{The setup}

To study the tachyon field non-minimally coupled to a Gauss-Bonnet term,  we consider the following action
\begin{align} \label{zz1}
S &=\int  d^4x \sqrt{-g} \left[ R + f({\phi}) R_{GB} - V(\phi) \sqrt{1+ \partial^{\alpha} \phi \partial_{\alpha} \phi}\right] + \int d^{4}x \mathcal{L}_{M}\left(g_{\mu\nu},\Psi_{M}\right),
\end{align}
where  $R$ is Ricci scalar, $g$ is determinant of the metric, $\phi$ is the tachyon scalar field, $\mathcal{L_{M}}$ represents the matter fields  and $R_{GB}$ is the Gauss-Bonnet curvature given by
\begin{equation}
R_{GB} = R^{2} - 4 R_{\mu \nu} R^{\mu \nu} + R_{\mu \nu \alpha \beta} R^{\mu\nu \alpha \beta},
\end{equation}
and we work in Planckian units where $\hbar = c = 8 \pi G =1$.
Variation of  action (\ref{zz1}) with respect to the metric gives the following field equation

\begin{align}\label{cfv}
G_{\mu\nu} = T^{(t)}_{\mu\nu}.
\end{align}
Here, $G_{\mu\nu}$ is the Einstein tensor and the total energy momentum tensor reads 

\begin{equation}
T^{(t)}_{\mu\nu}= T^{(\phi)}_{\mu\nu}+ T^{(\gamma)}_{\mu\nu},
\end{equation}
where $T^{(\gamma)}$ is the energy momentum tensor for radiation field and the energy momentum tensor for tachyon field can be expressed as \cite{g32, p1}

\begin{align}
\nonumber T^{(\phi)}_{\mu\nu}& = \frac{V(\phi) \partial_{\mu}\phi \partial_{\nu}\phi}{\sqrt{1+\partial^{\alpha}\phi \partial_{\alpha}\phi}}- g_{\mu\nu} V(\phi)\sqrt{1+\partial^{\alpha}\phi \partial_{\alpha}\phi}+ 2 \left[\nabla_{\mu} \nabla_{\nu} f(\phi) \right]R- 2 g_{\mu\nu}[\nabla^{2}f(\phi)]R - 4[\nabla_{\rho}\nabla_{\mu} f(\phi)]R^{\rho}_{\nu} \\& -4[\nabla_{\rho}\nabla_{\nu}f(\phi)]R^{\rho}_{\mu}+4[\nabla^{2}f(\phi)]R_{\mu\nu}+4 g_{\mu\nu}[\nabla_{\rho}\nabla_{\sigma}f(\phi)]R^{\rho\sigma}+ 4[\nabla^{\rho}\nabla^{\sigma}f(\phi)]R_{\mu\rho \sigma \nu}.
\end{align}
The equation of motion of the tachyon field is obtained by using Euler-Lagrange equation  \cite{p3}
\begin{align}\label{ss1}
-& \frac{V \ \Box \phi}{\left(1+ \nabla^{\alpha}\phi \nabla_{\alpha}\phi \right)^{\frac{1}{2}}} + \frac{V \  \nabla_{\mu}\nabla_{\nu}\phi}{\left(1+ \nabla^{\alpha}\phi \nabla_{\alpha}\phi\right)^{\frac{3}{2}}} \nabla^{\mu}\phi \nabla^{\nu}\phi - f^{\prime} R_{GB} + \frac{ V ^{\prime}}{{\left(1+ \nabla^{\alpha}\phi \nabla_{\alpha}\phi\right)^{\frac{1}{2}}}} =  0,
\end{align}
where setting $f = 0$, equation (\ref{ss1}) reduces to equation (4) in \cite{p3}. In the context of warm inflation, the inflaton field should decay into a radiation field and this is achieved in equation (\ref{ss1}) by adding a dissipation term $- \Gamma u^{\mu} \partial_{\mu} \phi$ to the right hand side \cite{g14, g15, g53}. Equation (\ref{ss1}) then takes the following form
\begin{align}\label{s1}
 & - \Box \phi + \frac{\nabla_{\mu}\nabla_{\nu}\phi}{1+ \nabla^{\alpha}\phi \nabla_{\alpha}\phi} \nabla^{\mu}\phi \nabla^{\nu}\phi - \frac{f^{\prime} R_{GB}}{V} \sqrt{1+ \nabla^{\alpha}\phi \nabla_{\alpha}\phi} + \frac{ V ^{\prime}}{V}= - \frac{\Gamma}{V} \left(\sqrt{1+ \nabla^{\alpha}\phi \nabla_{\alpha}\phi} \right)u^{\mu} \nabla_{\mu} \phi,
\end{align}
where a prime denotes derivation with respect to $\phi$, $u_{\mu}$ is the velocity four-vector with $u_{0}= -1$ and $\Gamma$ is a dissipation coefficient as a function of $\phi$. Since our model pertains to warm inflation, total energy momentum tensor contains both the inflaton and radiation fields with the inflaton field dominating over the radiation field at the beginning of inflation.

The symmetric energy momentum tensor can be uniquely decomposed according to fluids quantities as follows

\begin{equation}
 T^{(t)}_{\mu\nu}= \rho^{(t)} u_{\mu}u_{\nu}+P^{(t)} h_{\mu\nu}+ q^{(t)}_{\mu} u_{\nu}+ q^{(t)}_{\nu} u_{\mu} + \pi^{(t)} _{\mu\nu},
\end{equation}
where $h_{\mu\nu} = g_{\mu\nu}  + u_{\mu}u_{\nu}$ is the projection tensor with $h^{\mu}_{\mu} = (0, 1, 1, 1)$,  $\rho^{(t)}$, $P^{(t)}$, $q^{(t)}_{\mu}$ and $\pi^{(t)}_{\mu\nu}$ are the energy density, pressure, energy flux and anisotropic pressure respectively with $ u^{\mu}q^{(t)}_{\mu}$=0, $\pi^{(t)}_{\mu\nu}$= $\pi^{(t)}_{\nu\mu}$ and $u^{\mu}\pi^{(t)}_{\mu\nu}$=0. Now, using the projection tensor and velocity four-vector, $\rho^{(t)}$, $P^{(t)}$, $q^{(t)}_{\mu}$ and $\pi^{(t)}_{\mu\nu}$ are given by \cite{g49}

\begin{align}
& \rho^{(t)} = T^{(t)}_{\mu\nu} u^{\mu} u^{\nu}, \ \   P^{(t)}= \frac{1}{3} T^{(t)}_{\mu\nu} h^{\mu\nu},\ \  q^{(t)}_{\mu}= - T^{(t)}_{\rho \sigma} u^{\sigma} h^{\rho}_{\mu},\ \  \pi^{(t)}_{\mu\nu}= T^{(t)}_{\rho \sigma} h^{\rho}_{\mu} h^{\sigma}_{\nu}- P^{(t)} h_{\mu\nu}.
\end{align}
Since radiation field is a perfect fluid,  $T^{(\gamma)}_{\mu\nu}$  has no energy flux and anisotropic pressure components, $q^{(\gamma)}_{\mu}$=$\pi^{(\gamma)}_{\mu\nu}$=0. Although the energy flux and anisotropic pressure for radiation and tachyon fields are vanishing in a non-perturbative background, in the following section we will observe that such quantities are non-vanishing in a perturbative background.
Let us now proceed further by considering a spatially flat FRW metric

\begin{equation}
ds^{2} = - dt^{2} + a(t)^{2}\left(dx^{2}+ dy^{2} + dz^{2}\right),
\end{equation}
where $a(t)$ is the scale factor. Therefore, equation (\ref{cfv}) results in the following Friedmann equation

\begin{align}
H^{2} = \frac{1}{3}(\rho_{\phi}+ \rho_{\gamma}),
\end{align}
and the tachyon energy density and pressure can now be written as \cite{p1}

\begin{equation}
\rho_{\phi}= \frac{V(\phi)}{\sqrt{1-\dot \phi^{2}}}- 12 H^3 \dot f,
\end{equation}

\begin{align}
 P_{\phi}&= - V(\phi)\sqrt{1-\dot \phi^{2}}+ 4 H^{2}\ddot f+ 8 H \dot H \dot f + 8 H^{3} \dot f,
\end{align}
where a dot represents derivation with respect to the cosmic time and $q^{(\phi)}_{\mu}$=$\pi^{(\phi)}_{\mu\nu}$=0 for tachyon field energy momentum tensor and in fact it behaves like a perfect fluid. At the beginning of inflation $\rho_{\phi} \gg \rho_{\gamma} $ and therefore the Friedmann equation reduces to
\begin{equation}\label{zz2}
H^{2} \simeq \frac{1}{3} \rho_{\phi}.
\end{equation}

Let us now denote the radiation energy density by $\rho_{\gamma}$ with the equation of state give by  $P_{\gamma}=\frac{\rho_{\gamma}}{3}$.
In the warm inflationary model the inflaton field will decay to a radiation field at the end of inflation. The equation of motion now takes the form
\begin{equation}\label{d4}
\frac{\ddot \phi}{1- \dot \phi^{2}} + 3 H \dot \phi + \frac{V^{\prime}}{V} -  R_{GB} \frac{f^{\prime}}{V} \sqrt{1- \dot \phi^{2}} = - \frac{\Gamma}{V} \dot \phi \sqrt{1-\dot \phi^{2}},
\end{equation}
where $R_{GB} = 24 H^{2}\left(H^{2}+\dot H\right)$. The coupling between the Gauss-Bonnet curvature and tachyon field brings to the fore a new degree of freedom and following  \cite{g46}, one may define the hierarchy flow functions as follows
\begin{align}
&\epsilon_{1} = - \frac{\dot H}{H^{2}}, \ \  \ \ \ \ \ \ \ \ \epsilon_{i+1} = \frac{d \ln |\epsilon_{i}|}{d \ln a}, \\ & \delta_{1} = 4 H \dot f, \ \ \ \ \ \ \ \ \ \ \  \delta_{i+1} = \frac{d \ln |\delta_{i}|}{d \ln a},
\end{align}
where $i\ge 1$. In fact, we consider the Gauss-Bonnet coupling as not having any contribution to the energy density due to this new parameter and the standard slow-roll parameters become $|\epsilon_{i}| \ll 1$ and $|\delta_{i}| \ll 1$. During inflation we consider the slow-roll approximation where $\dot \phi^{2}\ll 1$ and $\ddot \phi \ll 3H\dot \phi$. Applying these approximations and the generalized slow-roll parameters, equation (\ref{zz2}) and (\ref{d4}) reduce to
\begin{equation}\label{sdf}
H^{2} \simeq \frac{V}{3},
\end{equation}

\begin{equation}\label{d2}
H \dot \phi \simeq - \frac{1}{3(1+D)}V Q,
\end{equation}
where $Q = \frac{V^{\prime}}{V^{2}} - \frac{8}{3} \ f^{\prime}$ and $D$ defines the dissipation factor  $D=\frac{\Gamma}{3 H V}$.
The Hubble and Gauss-Bonnet flow functions can now be expressed in general forms
\begin{align}\label{pp1}
\epsilon_{1} = \frac{Q}{2\left(1+D\right)}\frac{V^{\prime}}{V},
\end{align}

\begin{align}\label{pp2}
\epsilon_{2} = - \frac{Q}{\left(1+D\right)} \left( \frac{V^{\prime \prime}}{V^{\prime}} + \frac{Q^{\prime}}{Q} - \frac{V^{\prime}}{V} - \frac{D^{\prime}}{1+D} \right),
\end{align}

\begin{align}\label{pp3}
\delta_{1} = - \frac{4}{3\left(1+D\right)} f^{\prime} Q V,
\end{align}

\begin{align}\label{pp4}
\delta_{2} = - \frac{Q}{\left(1+D\right)} \left(\frac{f^{\prime \prime}}{f^{\prime}} + \frac{Q^{\prime}}{Q} + \frac{V^{\prime}}{V} - \frac{D^{\prime}}{1+D}\right),
\end{align}
where a prime denotes derivation with respect to the field. We note that inflation takes place for $\epsilon_{1} < 1$ and terminates when $\epsilon_{1} \simeq 1$. The e-folding  should now be calculated as  the criteria for a viable inflation. In our model the e-folding can be calculated as a function of $\phi$
\begin{equation}\label{pp5}
N(\phi) \equiv \int^{t_{end}}_{t_{hc}} H dt = \int^{\phi_{end}}_{\phi_{hc}} \frac{H}{\dot \phi} d \phi \simeq \int^{\phi_{hc}}_{\phi_{end}} \frac{\left(1+D\right)}{Q} d \phi,
\end{equation}
where $\phi_{hc}$ and $\phi_{end}$ denote the values of the scalar field at the Hubble crossing time and termination of inflation. The conservation equations for both radiation and inflaton is given by
\begin{equation}
\dot \rho ^{(t)} + 3 H \left(\rho^{(t)}+ P^{(t)}\right) = 0.
\end{equation}
The energy density and pressure of the radiation field can be related to entropy \cite{g52}
\begin{equation}
\rho^{(t)} = \rho_{\phi} + \rho_{\gamma}= \rho_{\phi}+ \frac{3}{4} ST,
\end{equation}

\begin{equation}
 P^{(t)} = P_{\phi}+P_{\gamma} = P_{\phi}+ \frac{1}{4} ST,
\end{equation}
where $T$ and $S$ denote temperature and entropy respectively. The conservation equation for a tachyon field in the presence of dissipation takes the form
\begin{equation}
\dot \rho_{\phi} + 3 H\left(\rho_{\phi}+ P_{\phi}\right)= - \Gamma \dot \phi^{2}.
\end{equation}
One may then write the entropy production for radiation field during the inflationary phase
\begin{equation}
 T\left(\dot S+ 3H S\right)= \Gamma \dot \phi^{2},
\end{equation}
using the conservation equation
\begin{equation}
\dot \rho_{\gamma} + 4 H \rho_{\gamma} = \Gamma \dot \phi^{2}.
\end{equation}

During inflation the radiation field production can be considered as quasi-stable so that $\dot \rho_{\gamma}\ll 4H\rho_{\gamma}$ and  $\dot \rho_{\gamma}\ll \Gamma \dot \phi^{2}$, therefore
\begin{equation}\label{zz3}
\rho_{\gamma} = \frac{\Gamma \dot \phi^{2}}{4 H}= \sigma T^{4}_{r},
\end{equation}
where $\sigma $ is the Boltzman constant. The relation between energy density of radiation and the inflaton field can be calculated using the slow-roll parameter
\begin{equation}\label{zz4}
\rho_{\gamma} =\epsilon_{1} \frac{\Gamma Q H \rho_{\phi}}{2\left(1+D\right)\rho_{\phi}^{\prime}},
\end{equation}
where $\rho_{\phi} \simeq V$. Using the condition for inflation, $\epsilon_{1} \ < 1$, from the above equation we have
\begin{equation}
\rho_{\gamma} <  \frac{\Gamma Q H \rho_{\phi}}{2\left(1+D\right)\rho_{\phi}^{\prime}} .
\end{equation}
This condition will exist during the inflationary period.

                                                                                 \section{Perturbations}

In this section we study perturbations of the FRW background in the longitudinal gauge and present a complete set of perturbed equations. We begin by writing the perturbed FRW metric
\begin{equation}
ds^{2} = - \left(1+ 2 \Phi \right) dt^{2} + a(t)^{2} \left(1-2 \Psi \right) \delta_{ij} dx^{i} dx^{j},
\end{equation}
where $\Phi$ and $\Psi$ are gauge invariant metric perturbation quantities. The spatial dependence of all perturbed quantities are of the form of a plane wave $e^{ik.x}$, where $k$  is the wave number. All perturbed equations in Fourier space for matter can be calculated with the result \cite{g49}
\begin{align}
&-6H\left(H\Phi + \dot \Psi\right)-2 \frac{k^{2}}{a^{2}} \Psi = \delta \rho^{(t)},\\ &
 2 \ddot \Psi + 4 \dot H\Phi + 2 H \dot \Phi+ 6 H \dot \Psi + 6 H^{2} \Phi+ \frac{k^{2}}{a^{2}}\left(\Psi- \Phi\right)  =  \delta P^{(t)},\\&
- 2 \left(H \Phi+ \dot \Psi\right) = \delta q^{(t)}_{j},\\&
-\frac{1}{a^{2}}(\Phi-\Psi)^{|i}_{|j}= \delta \pi^{i\ (t)}_{j}.
\end{align}
Here the perturbed matter quantities contain both radiation and inflaton fields
\begin{align}
 &\delta \rho^{(t)} = \delta \rho_{\phi} + \delta \rho_{\gamma} ,\\& \delta P^{(t)} = \delta P_{\phi} + \delta P_{\gamma},\\&  \delta q^{(t)}_{j}= \delta q^{\phi}_{j}+\delta q^{\gamma}_{j}= \delta q^{(\phi)}_{j}- \left(\rho_{\gamma}+P_{\gamma}\right) \delta u_{j}, \\ &  \delta \pi^{(t)}_{ij}=\delta \pi^{\phi}_{ij}+\delta \pi^{\gamma}_{ij}.
\end{align}

The perturbed conservation equations for the radiation field are \cite{l2}
\begin{equation}\label{d6}
\delta \dot \rho_{\gamma}+ 4 H \delta \rho_{\gamma} + \frac{4}{3} \frac{k}{a} \rho_{\gamma} \nu = 4 \rho_{\gamma} \dot \Psi + 2 \Gamma \dot \phi \delta \dot \phi - \Gamma \dot \phi^{2} \Phi + \Gamma^{\prime} \dot \phi^{2} \delta \phi,
\end{equation}
\begin{equation}\label{d7}
\dot \nu + 4 H \nu+ \frac{k}{a}\left[ \Phi+ \frac{\delta \rho_{\gamma}}{4 \rho_{\gamma}}+ \frac{3 \Gamma \dot \phi}{4 \rho_{\gamma}} \delta \phi\right]=0,
\end{equation}
where $\delta u_{i}$ decomposes as $\delta u_{j} = - \frac{i a k_{j}}{k} \nu e^{i k x}$ $\left(j=1, 2, 3\right)$ \cite{l1} and here we have omitted subscript $k$, with the perturbed quantities of the field taking the form
\begin{align}
\delta \rho_{\phi}&=\frac{V^{\prime} \delta \phi}{\sqrt{1- \dot \phi^{2}}}+ \frac{V \dot \phi \delta \dot \phi - V\dot \phi^{2} \Phi}{\left(1-\dot \phi^{2}\right)^{\frac{3}{2}}}- 12 H^{3} \left(f^{\prime \prime} \dot \phi \delta \phi+ f^{\prime }\delta \dot \phi\right)+12 H^{2} f^{\prime}\dot \phi \left(4H \Phi+3\dot \Psi \right)+\frac{4H k^{2}}{a^{2}}\left(2f^{\prime}\dot \phi \Psi - Hf^{\prime}\delta \phi\right),
\end{align}

\begin{align}
\nonumber \delta P_{\phi}= &- V^{\prime} \sqrt{1- \dot \phi^{2}} \delta \phi + \frac{V \dot \phi \delta \dot \phi-V \dot \phi^{2} \Phi}{1-\dot \phi^{2}}+4 H^{2} \delta \ddot f - 32 H \Phi\left[H\left(f^{\prime \prime} \dot \phi^{2}+f^{\prime} \ddot \phi \right)+\dot H f^{\prime} \dot \phi+ H^{2} f^{\prime} \dot \phi\right]+ 4 H^{2} \dot \Phi \left[4 \left(f^{\prime \prime} \dot \phi^{2} \right. \right. \\& \left. \left. \nonumber +f^{\prime} \ddot \phi \right)-3 f^{\prime }\dot \phi \right]- 8 \dot \Psi \left[H \left(f^{\prime \prime} \dot \phi^{2}+f^{\prime} \ddot \phi \right)+ \dot H f^{\prime} \dot \phi+6 H^{2} f^{\prime} \dot \phi\right] - 8 H f^{\prime} \dot \phi \ddot \Psi+\left(8H \dot H- 8H^{3}\right)\left(f^{\prime \prime} \dot \phi \delta \phi + f^{\prime} \delta \dot \phi\right)\\& +\left(4 \dot H  -4 H^{2}\right) \frac{k^{2}}{a^{2}}\left(f^{\prime}\delta \phi\right)+4 \frac{k^{2}}{a^{2}} \left[Hf^{\prime}\dot \phi \Phi- \left(f^{\prime \prime} \dot \phi^{2}+f^{\prime} \ddot \phi \right)\Psi\right],
\end{align}

\begin{align}\label{d5}
 \delta q^{\phi}_{j}=-\frac{V \dot \phi \delta \phi}{\sqrt{1- \dot \phi^{2}}} - 4 H^{2}\left(f^{\prime \prime} \dot \phi \delta \phi + f^{\prime} \delta \dot \phi\right) + 12 H^{2} f^{\prime} \dot \phi \Phi + 4 H^{3} f^{\prime} \delta \phi + 8 H f^{\prime} \dot \phi \dot \Psi,
\end{align}

\begin{equation}
 \delta \pi^{i (\phi)}_{j} =\frac{1}{a^{2}} \left[-4 \left(f^{\prime \prime }\dot \phi^{2}+ f^{\prime} \ddot \phi \right) + 4 H f^{\prime} \dot \phi \Phi + 4 \left(H^{2}+\dot H \right) f^{\prime} \delta \phi \right]^{|i}_{|j}.
\end{equation}

We can also obtain  perturbed equation of motion for the tachyon field  using equation (\ref{s1})
\begin{align}\label{d8}
\nonumber & \frac{\delta \ddot \phi-\dot \phi \dot \Phi- 2 \ddot \phi \Phi + 2 \ddot \phi \dot \phi \delta \dot \phi- 2 \dot \phi ^{2} \ddot \phi \Phi}{1- \dot \phi^{2}}+3H \delta \dot\phi - 3 \dot \phi \dot \Psi - 6 H \dot \phi \Phi+\frac{k^{2}}{a^{2}} \delta \phi+\frac{\ddot \phi \dot \phi ^{3}\delta \dot \phi-\ddot \phi \dot \phi^{4}\Phi}{\left(1-\dot \phi^{2}\right)^{2}}+\left(\frac{V^{\prime \prime}}{V}-\left(\frac{V^{\prime}}{V}\right)^{2}\right)\delta \phi \\&-\frac{8}{3} \left[ V^{\prime} f^{\prime} + V f^{\prime \prime} \right] \delta \phi \ \sqrt{1-\dot \phi ^{2}}+ \frac{8}{3}f^{\prime} V \left(\frac{\dot \phi \delta \dot \phi-\dot \phi^{2} \Phi}{\sqrt{1-\dot \phi^{2}}}\right) = \left[\frac{\Gamma}{V}\dot \phi \Phi - \frac{\Gamma}{V} \delta \dot \phi- \frac{\Gamma^{\prime}}{V} \dot \phi \delta \phi+ \frac{V^{\prime}}{V^{2}} \Gamma \dot \phi \delta \phi \right] \sqrt{1 -\dot \phi^{2}}\nonumber \\&+ \frac{\Gamma}{V}\frac{ \dot \phi^{2} \delta \dot \phi}{\sqrt{1 - \dot \phi^{2}}} - \frac{\Gamma}{V} \frac{\dot \phi^{3} \Phi}{\sqrt{1-\dot \phi^2}}.
\end{align}
The above equations describe dynamics of our inflationary model and the parameters of interest can be calculated using them.

                                                                                 \section{The Power spectrum}

In the previous section we obtained a complete set of perturbed equations, which, due to their complexity cannot be solved in the presence of higher order time-derivative of perturbed quantities. In \cite{g50} and \cite{g51}  it is shown that during inflation one may consider perturbed quantities as changing slowly which makes it plausible to neglect $\dot \Phi$, $\dot \Psi$ and $\ddot \Psi$. In fact, for the longitudinal post-Newtonian limit to be satisfied, we require that $\Delta \Psi \gg a^{2} H^{2}\times(\Psi, \dot \Psi, \ddot \Psi)$ and similarly for other gradient terms. For a plane wave perturbation with wavelength $\lambda$, we see that $H^{2}\Psi$ is much smaller than $\Delta \Psi$ when $\lambda \ll \frac{1}{H}$. The requirement that $\dot \Psi$ be also negligible implies the condition $\frac{d \log \Psi}{d \xi}\ll \frac{1}{(\lambda H)^{2}}$, with $\xi = \log a$, which holds if the condition $\lambda \ll \frac{1}{H}$ is satisfied for perturbation growth. This argument can be applied to $\ddot \Psi$ and the other metric potential, namely $\Phi$ too.  Now, using equation (\ref{d4}) and the slow-roll conditions, equations (\ref{d6}, \ref{d7}, \ref{d5}, \ref{d8}) reduce to
\begin{align}\label{f1}
\left( 3H + \frac{\Gamma}{V}\right) \delta \dot \phi + \left[ \frac{V^{\prime \prime}}{V}-\left (\frac{V^{\prime}}{V}\right)^{2}-\frac{8}{3} V^{\prime} f^{\prime}-\frac{8}{3} V f^{\prime \prime} + \frac{\Gamma^{\prime}}{V} \dot \phi  - \frac{V^{\prime}}{V^{2}}\Gamma \dot \phi \right] \delta \phi + \left[ -\frac{\Gamma}{V} \dot \phi + \frac{2V^{\prime}}{V} - \frac{16}{3} V f^{\prime} \right] \Phi \simeq 0,
\end{align}

\begin{align}\label{f2}
 2H \Phi & \simeq \left[ -\frac{4}{3}\frac{\rho_{\gamma}a \nu}{k} + V \dot \phi \delta \phi - 4 H^{3} f^{\prime} \delta \phi \right],
\end{align}

\begin{equation}\label{f6}
\nu \simeq -\frac{k}{4 a H} \left[\Phi + \frac{\delta \rho _{\gamma}}{4 \rho_{\gamma}}+ \frac{3 \Gamma \dot \phi}{ 4 \rho_{\gamma}} \delta \phi\right],
\end{equation}

\begin{equation}\label{f5}
\frac{\delta \rho_{\gamma}}{\rho_{\gamma}} \simeq \frac{\Gamma^{\prime}}{\Gamma} \delta \phi - \Phi,
\end{equation}
where we can rewrite equation (\ref{f2}) using equations (\ref{f6},\ref{f5})
\begin{equation}\label{s3}
\Phi \simeq \frac{V \dot \phi \delta \phi}{2 H}\left[1+ \frac{\Gamma}{4H V}+\frac{\Gamma^{\prime} \dot \phi}{48H^{2}V} -\frac{4H^{3}f^{\prime}}{V\dot \phi}\right].
\end{equation}
Equation (\ref{f1}) can be solved by taking into account the tachyon field as an independent variable in place of the cosmic time \cite{g39}. Using equation (\ref{d2}) we have the following
\begin{align}
& \left(3H+\frac{\Gamma}{V}\right)\frac{d}{dt} = \left(3H+\frac{\Gamma}{V}\right) \dot \phi \frac{d}{d\phi}= - \left(\frac{V^{\prime}}{V} - \frac{8}{3} V f^{\prime}\right) \frac{d}{d \phi}.
\end{align}
Equation (\ref{f1}) can then be rewritten as a first order differential equation with respect to $\phi$
\begin{align}
\nonumber\frac{(\delta \phi)^{\prime}}{\delta \phi}&=\frac{1}{\left(\frac{V^{\prime}}{V} - \frac{8}{3} V f^{\prime} \right)}\left(\left(\frac{V^{\prime}}{V} - \frac{8}{3} V f^{\prime} \right)^{\prime} + \left(\frac{\Gamma}{V}\right)^{\prime}\dot \phi  +\left(-\frac{\Gamma}{V} \dot \phi + \frac{2V^{\prime}}{V} - \frac{16}{3} V f^{\prime} \right) \left(\frac{V \dot \phi}{2 H}\right) \right. \\&\left. \times\left[1+ \frac{\Gamma}{4H V}+\frac{\Gamma^{\prime} \dot \phi}{48H^{2}V} -\frac{4H^{3}f^{\prime}}{V\dot \phi}\right]\right).
\end{align}
Now, following {\cite{g39,l1,g54,g55}  we define the auxiliary function
\begin{equation}
\chi(\phi) \equiv \frac{\delta \phi}{\left(\frac{V^{\prime}}{V} - \frac{8}{3} V f^{\prime}\right)} \exp\left(\int \frac{\left(\frac{\Gamma}{V}\right)^{\prime}}{\left(3H+\frac{\Gamma}{V}\right)}d \phi\right).
\end{equation}
Using the above definition we find
\begin{equation}
\frac{\chi^{\prime}(\phi)}{\chi(\phi)}= \frac{\left(\delta \phi\right)^{\prime}}{\left(\delta \phi\right)} - \frac{\left(\frac{V^{\prime}}{V} - \frac{8}{3} V f^{\prime}\right)^{\prime}}{\left(\frac{V^{\prime}}{V} - \frac{8}{3} V f^{\prime}\right)}+\frac{\left(\frac{\Gamma}{V}\right)^{\prime}}{\left(3H+\frac{\Gamma}{V}\right)}.
\end{equation}
We may now calculate the following expression
\begin{align}
 & \frac{\chi^{\prime}(\phi)}{\chi(\phi)}=\left(-\frac{\Gamma}{V} \dot \phi + \frac{2V^{\prime}}{V} - \frac{16}{3} V f^{\prime} \right)\left(\frac{V \dot \phi}{2 H}\right) \times\left(1+ \frac{\Gamma}{4H V}+\frac{\Gamma^{\prime} \dot \phi}{48H^{2}V} -\frac{4H^{3}f^{\prime}}{V\dot \phi}\right).
\end{align}
This equation has an explicit solution given by
\begin{align}
\chi(\phi)&= C \exp \left(- \int \frac{9}{8} \frac{2H + \frac{\Gamma}{V}}{\left(3H + \frac{\Gamma}{V}\right)^{2}}\left[\Gamma + 4 H V - \frac{\Gamma^{\prime}V Q}{12H\left(3H + \frac{\Gamma}{V}\right)} - \frac{48\left(1+D\right) H^{5} f^{\prime}}{V Q}\right]Q \right),
\end{align}
where $C$ is an integration constant and  $\delta \phi$ can be written as
\begin{equation}\label{s2}
\delta \phi = C\left(\frac{V^{\prime}}{V} - \frac{8}{3} V f^{\prime}\right) \exp \left(\zeta (\phi)\right),
\end{equation}
where
\begin{align}\label{fff}
 \zeta(\phi)&= \exp \left(-\int \frac{(\frac{\Gamma}{V})^{\prime}}{3H+\frac{\Gamma}{V}} + \frac{9}{8} \frac{2H + \frac{\Gamma}{V}}{\left(3H + \frac{\Gamma}{V}\right)^{2}}\left[\Gamma + 4 H V - \frac{\Gamma^{\prime}V Q}{12H\left(3H + \frac{\Gamma}{V}\right)} - \frac{48\left(1+D\right) H^{5} f^{\prime}}{V Q}\right] Q\right).
\end{align}
The density perturbation is then \cite{q9,g57, g39}
\begin{equation}\label{s4}
\delta_{H}\equiv \frac{2}{5} p^{\frac{1}{2}}_{\mathcal{R}}= \frac{16 \pi}{5} \frac{\exp(- \zeta(\phi))}{\left(\frac{V^{\prime}}{V} - \frac{8}{3} V f^{\prime}\right)} \delta \phi.
\end{equation}
In fact the second term in the denominator results from the Gauss-Bonnet modification which upon setting  $f=0$, equations (\ref{fff}) and (\ref{s4})  reduce to equation (31) and (32) in \cite{g39} and  the amplitude of curvature perturbation for $\Gamma = 0$ and $f=0$ goes to $\delta_{H} \simeq \frac{H}{\dot \phi} \delta \phi$, corresponding to cold inflation. The above equation would enable us to obtain the spectral index and its running. The aim of the next section is to investigate the model in the high dissipation regime in order to obtain the general form of the modified spectral index and tensor-to-scalar ratio.

                                                                     \section{High dissipation regime ($D\gg 1$)}

To achieve what just mentioned above, we note that in warm inflationary models the fluctuation of the scalar field in a high dissipative regime may be generated by thermal fluctuations instead of quantum fluctuations. This then means that \cite{g61}
\begin{equation}
(\delta \phi)^{2} \simeq \frac{k_{F} T_{r}}{2 \pi^{2}},
\end{equation}
where in this limit for the frozen-out wave number we have $k_{F}= \sqrt{\frac{\Gamma H}{V}}=H \sqrt{3D}\ge H $.
With the help of this equation in high dissipation regime ($D\gg1$) we find
\begin{equation}
{\delta}_{H}^{2} = \frac{64\sqrt{3}}{75 }T_{r} \frac{{V}^{\prime} exp (- 2{\zeta}(\phi))}{\sqrt{{D}} {\epsilon}_{1} {Q}V^2 {H}}.
\end{equation}
Using (\ref{zz3}, \ref{zz4}) one can also obtain
\begin{equation}
{\delta}_{H}^{2} = \frac{64}{75}\left(\frac{27}{2 \sigma}\right)^{\frac{1}{4}}\left(\frac{ { {V^{\prime}}}^{3}}{{{D}^{2}} {\epsilon}_{1}^{3} {Q}^{3} V^{6} { H}^{2}}\right)^{\frac{1}{4}} \exp (- 2{\zeta}(\phi)),
\end{equation}
where
\begin{align}\label{gg1}
{\zeta} (\phi)&= \exp \left(-\int \frac{(\frac{{\Gamma}}{V})^{\prime}}{3 {H} {D}} + \frac{9}{8} \left[1 - \frac{\left(\ln {\Gamma} \right)^{\prime}V {Q}}{36{H}^{2} {D}}  - \frac{48\left({D}\right) {H}^{5} f^{\prime}}{V {Q} {\Gamma}}\right] {Q} V\right) .
\end{align}

One of the most important parameters to consider is the scalar spectral index which can be obtained as follows
\begin{align}
{n}_{s}=1+\frac{d \ln \delta_{H}^{2}}{d \ln k} = 1+\frac{d \ln \delta_{H}^{2}}{d \ln N}= 1 +\frac{d \ln \delta_{H}^{2}}{d \phi} \frac{d \phi}{d{N}}.
\end{align}
It can also be expressed in terms of generalized slow-roll parameters
\begin{align}\label{pp6}
{n}_{s}& = 1+\frac{13}{2} {\epsilon}_{1} -  \frac{1}{4} {\epsilon}_{2} - 3 {\delta}_{1} - {\epsilon}_{1} Z(\phi),
\end{align}
where use has been made of the following equation
\begin{align}
{\zeta}^{\prime} (\phi)= -\left(\ln(\frac{{\Gamma}}{V})\right)^{\prime} - \frac{9}{8} \left[QV- \frac{\left(\ln {\Gamma} \right)^{\prime}{Q}^{2} V}{12 {D}}+ \frac{4 {D}}{3 Q} {\delta}_{1}\right],
\end{align}
and
\begin{align}
Z(\phi) \equiv \left(\frac{V}{V^{\prime}}\right)\left(\frac{9}{2}Q V+ 4 \left(\ln \Gamma\right)^{\prime} - \frac{5}{2} \frac{Q^{\prime}}{Q}+ \frac{1}{2}\frac{V^{\prime \prime}}{V^{\prime}}\right),
\end{align}
and the fact that $d \ln k \simeq d N(\phi)$ \cite{g9}. We may also obtain the running index
\begin{align}\label{mm1}
{\alpha}_{s}= \frac{d {n}_{s}}{d\ln k} = \frac{d {n}_{s}}{d\phi} \frac{d\phi}{d {N}}= \frac{{n}_{s}^{\prime}}{{N}^{\prime}}.
\end{align}
This can be written in the terms of the slow-roll parameters as follows
\begin{align}\label{mm2}
 {\alpha}_{s} = \frac{13}{2} {\epsilon}_{1} {\epsilon}_{2} - \frac{1}{4} {\epsilon}_{2} {\epsilon}_{3} - 3 {\delta}_{1} {\delta}_{2} - {\epsilon}_{1} {\epsilon}_{2}  Z(\phi) + 2 {\epsilon}_{1}^2 \left(\frac{V}{V^{\prime}}\right) Z^{\prime}(\phi).
\end{align}
The amplitude for tensor perturbation is given by
\begin{equation}
A_{t} = 2 \left(\frac{H}{2 \pi}\right)^{2} \coth \left(\frac{k}{2T}\right) = \frac{V}{6 \pi^{2}} \coth \left(\frac{k}{2T}\right),
\end{equation}
where $T$ is the thermal background of gravitational waves \cite{g63}.  Using the above equation, one  obtains the spectral index for gravitational waves
\begin{equation}\label{mm3}
n_{t} = \frac{d \ln \left(\frac{A_{t}}{\coth \left(\frac{k}{2T}\right) }\right)}{d \ln k} \simeq - 2 \epsilon_{1}.
\end{equation}
We may also derive the tensor-to-scalar ratio
\begin{align}\label{mm4}
&r(k_{0}) = \frac{A_{t}}{\mathcal{P}_{r}}\Big|_{k=k_{0}}= \frac{{H}^{2}}{32 \pi^{2}} \left(6 \sigma\right)^{\frac{1}{4}}\left(\frac{{{D}^{2}} {\epsilon}_{1}^{3} {Q}^{3} V^{6} { H^{2}}}{ { {V^{\prime}}}^{3}}\right)^{\frac{1}{4}} \exp (2{\zeta}(\phi)) \coth \left(\frac{k_{0}}{2T}\right),
\end{align}
where $\mathcal{P}_{r} = \frac{25}{4} \delta^{2}_{H}$ and $k_{0}$ denotes the value of $k$ when the scale of the universe crosses the Hubble horizon. An upper bound for tensor-to-scalar ratio is obtained using Planck 2015 data, $r< 0.12$ \cite{n1}.

                                                                                   \section{Exponential Potential}

In the following two sections, we will consider the relevant potentials and Gauss-Bonnet functions and integrate equation (\ref{pp5}) to find the value of the scalar field at the beginning of inflation in terms of the e-folding number $N$ in order to obtain analytical solutions and investigate predictions of the model. To this end, we take the potential and Gauss-Bonnet coupling functions as follows
\begin{equation}
V(\phi) = V_{0} \e ^{- \alpha \phi}, \ \ \ \ \ \ \ \ \ f(\phi) = \xi_{0} \e ^{\alpha \phi},
\end{equation}
where $V_{0}$, $\xi_{0}$ and $\alpha$ are constant. Although the principal role of a dissipative coefficient in a warm inflationary scenario has phenomenologically been studied over the past few decades, its exact functional form is still controversial. In this sense, many parameterizations have been introduced in order to treat the functional form of $\Gamma$. The simplest form is a constant dissipative coefficient, although it may have a general functional form of temperature $T$ and scalar field $\phi$  inspired from supersymmetry \cite{ccv}. On the other hand, it can be proportional to a potential ($\Gamma \propto V$) which has also been considered  \cite{g35, g39} in numerous works. The interested reader can find a full set of derivation of such an assumption in Appendix B of \cite{adf1} where the authors have invoked the method presented in \cite{adf2} which is based on  thermal effective field theory analysing intermediate particle production to achieve their goal. Therefore, taking $\Gamma \propto V$ one can rewrite the Hubble and Gauss-Bonnet flow parameters in terms of the e-folding using equations (\ref{pp1}, \ref{pp2}, \ref{pp3}, \ref{pp4}, \ref{pp5}) with the result
\begin{align}
{\epsilon}_{1} = {\epsilon}_{2} = {\delta}_{2} =  \frac{1}{N+1}, \ \ \ \ \  \ {\delta}_{1} = \frac{\beta}{N+1},
\end{align}
where we have defined $\beta = \frac{8}{3} V_{0} \xi_{0}$ for simplicity and $\epsilon_{1}$ is an increasing function for $\beta <-1$. Using the above equation and equations (\ref{pp6}, \ref{mm2}), we may write the spectral index and its running in the term of the e-folding number
\begin{align}\label{qq1}
n_{s}(\beta, N) = 1 - \frac{W}{N+1},
\end{align}
\begin{align}\label{bb1}
\alpha_{s}(\beta, N) = - \frac{W}{(N+1)^2},
\end{align}
with
\begin{align}
W \equiv \left(\frac{21}{4}+\frac{15}{2}\beta\right),
\end{align}
where the spectral index changes with the inverse e-folding which means that at large e-foldings it is scale invariant, as one would expect. From equations (\ref{qq1}, \ref{bb1}),  consistency with the released data suggests a new upper bound for $\beta$, namely $\beta< - 0 .7$  since $\beta > - 0.7$ results in $n_{s}> 1$. We may also conveniently write the spectral index for gravitational waves in terms of the e-folding
\begin{align}
n_{t}(N) = -\frac{2}{N+1}.
\end{align}
We are also able to express tensor-to-scalar ratio in the terms of $N$
\begin{align}
r(\alpha, \beta, V_{0}, \Gamma_{0},N) &= J \left(N+1\right)^{-\left(\frac{13}{4} + \frac{15}{2} \beta \right)} \exp \left(\frac{\frac{9}{4} + \frac{9}{4} \beta}{N+1}\right) \coth \left(\frac{k_{0}}{2 T}\right),
\end{align}
where
\begin{align}
J \equiv \frac{\alpha^{4}}{128 \pi^{2}} \left(\frac{2 \sigma}{3 \Gamma_{0}^{6}}(1+\beta)^{11} \right)^{\frac{1}{4}} \left(\frac{4 \Gamma_{0}^{2} V_{0}}{3 (1+\beta)^{2} \alpha^{4}}\right)^{\left(\frac{9}{4}+\frac{15}{4}\beta \right)},
\end{align}
\begin{figure}[H]
 \includegraphics[scale=0.7]{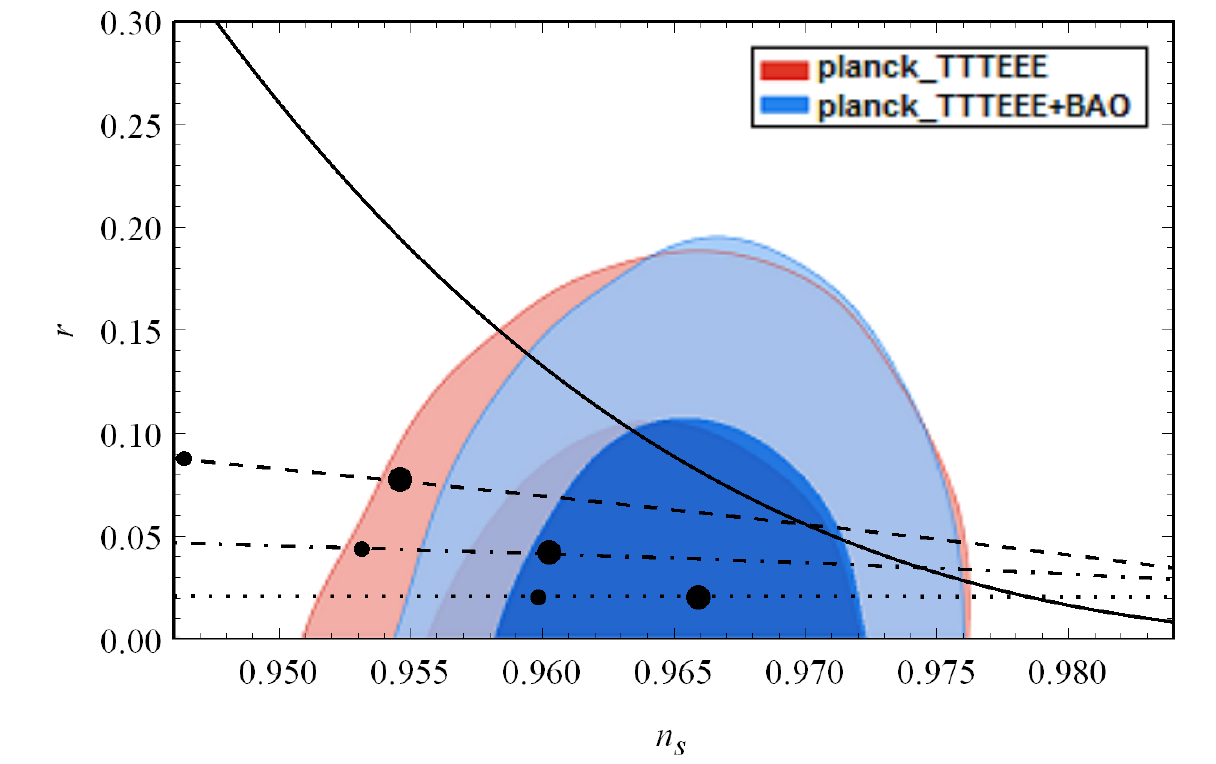} \ \ \ \ \ \ \ \ \includegraphics[scale=0.7]{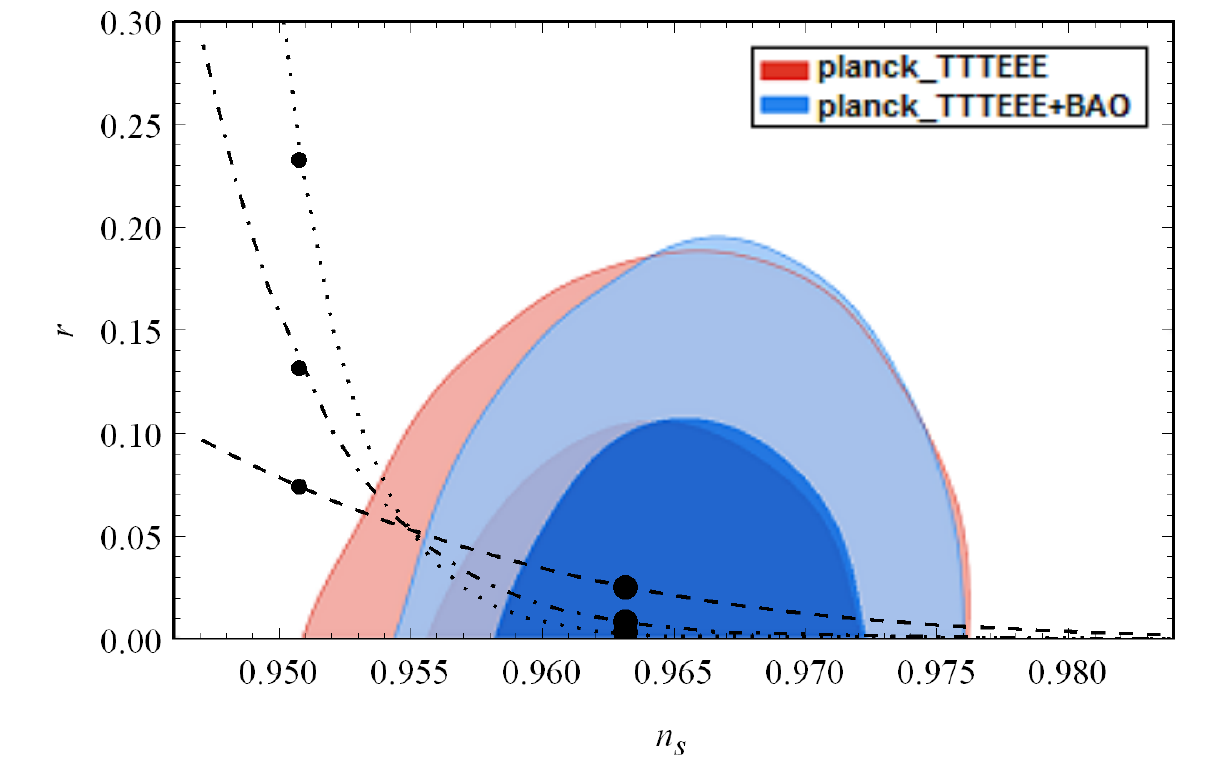}
\caption{Two-dimensional joint marginalized constraint (68\% and 95\% confidence level) on the scalar spectral index $n_{s}$ and tensor-to-scalar ratio, $ r$, including theoretical predictions where in the left panel, dashed, dot-dashed and dotted curves denote $\beta  = - 0.3$, $\beta =- 0.35$ and $\beta = -0.4$ respectively and small and large points represent N = 55  and N = 65 with $\Gamma_{0} = 150$ and $\alpha = 1$. The solid curve represents the model without Gauss-Bonnet coupling constant which is out of the panel for $N=55$ and $N=65$. Moreover, the value of e-folding increases moving from small to large points on each curve. In the right panel, dashed, dot-dashed and dotted curves denote $\Gamma_{0} = 100$, $\Gamma_{0} = 1000$ and $\Gamma_{0} = 10000$ respectively and small and large points represent $\beta = - 0.3$ and $\beta = - 0.4$ with N = 60 and $\alpha = 1$. In addition, the value of $\beta$ decreases from small to large points on each curve.}
\end{figure}
and $\Gamma_{0}$ is the amplitude of dissipation coefficient. In addition, during this study we have assumed that thermal background temperature of gravitational waves is equal to radiation field temperature. This means $T = T_{r}=\left(\frac{3}{8} \frac{(1+\beta)^{3} (N+1) \alpha^{4}}{\sigma\Gamma_{0}^{2}}\right)^{\frac{1}{4}} $ for the selected type of functions. Equations (\ref{qq1}, \ref{bb1}) are clearly showing that decreasing $\beta$ will result in enhancing the spectral index and its running. An interesting point is that the spectral index and its running for this type of functions are independent of dissipation coefficient amplitude $\Gamma_{0}$ and $\alpha$, meaning that these quantities shift the tensor-to-scalar ratio, $r$, and do not change the spectral index and its running. Therefore, they can control the value of tensor-to-scalar ratio in such a way as to be consistent with planck 2015 data. Using equation (\ref{qq1}), we  show the running of the spectral index, tensor-to-scalar ratio and spectral index of gravitational waves in  terms of the spectral index in order to have a better understanding of their behavior as follows
\begin{align}\label{hh3}
\alpha_{s}& = - W^{-1} (n_{s} -1)^{2},\\ n_{t} &= 2 W^{-1} (n_{s} -1),
\end{align}
\begin{widetext}
\begin{align}\label{hh4}
 r(\alpha, \beta, V_{0}, \Gamma_{0}, N) &= J \left(- \frac{(n_{s}-1)}{W}\right)^{\left( \frac{13}{4} +\frac{15}{2} \beta\right)} \exp \left( -\left(\frac{\frac{9}{4} + \frac{9}{4} \beta}{W}\right) (n_{s}-1)\right)  \coth \left(\frac{k_{0}}{2 T}\right),
\end{align}
\end{widetext}
\begin{align}
T=\left(\frac{3}{8} \frac{(1+\beta)^{3} (- \frac{W}{(n_{s} - 1)} ) \alpha^{4}}{\sigma\Gamma_{0}^{2}}\right)^{\frac{1}{4}}.
\end{align}
Use of relations (\ref{hh3}, \ref{hh4}) would enable us to compare our theoretical predictions with a two-dimensional joint marginalized constraint. The left panel in figure 1 shows three different values of $\beta$ and variation of the e-folding where for a fixed value of the e-folding, decreasing  $\beta$  horizontally shifts the spectral index and also  shifts $r$ vertically. In fact, decreasing the value of $\beta$ causes an enhancement in the spectral index and tensor-to-scalar ratio. As can be seen  in figure 1,  setting the coupling constant to zero is not in agreement with Planck data for fixed values of parameters of the model and is even out of the panel in the figure. In fact, it needs a large e-folding number, e.g. $N=120$, or even larger values to be within the $95\%$ region which is not reasonable. In the right panel of figure 1, we show the behavior of tensor-to-scalar ratio versus spectral index for variation of $\beta$ and three different values of $\Gamma_{0}$. This figure shows that our theoretical predictions for the behavior of the spectral index versus tensor-to-scalar ratio is divided into two regimes, strong and weak, for fixed values of $\beta$. For large fixed values of $\beta$, increasing the value of $\Gamma_{0}$ vertically shifts tensor-to-scalar ratio towards smaller values of $r$ and do not change the value of $n_{s}$ and this will  inversely happen for small fixed values of $\beta$. In fact, large $\beta$ results in a notable enhancement for $r$ by increasing $\Gamma_{0}$ and this will go outside the two-dimensional joint marginalized constraint by increasing $\Gamma_{0}$ for small values of $\beta$. It should be noted that in this section and the next we have taken $V_{0}$ =1  for plotting the figures and thus $\beta = \frac{8}{3} \xi_{0}$.

Finally, we have also attempted to reduce the number of parameters of the model using space parameters $n_{s}, \alpha_{s}, r$ and $n_{t}$ in order to find tighter constraints on the parameters. Here, since $\alpha_{s}, n_{s}$ and $n_{t}$ are independent of $\Gamma_{0}$ and $\alpha$  we  just need two equations to constrain our model. Using equations (\ref{qq1}, \ref{bb1}) one  finds the constraints on $\beta$ which are summarized in table I.

\begin{table}
\begin{tabular}{ |c|c|c|c| }
\hline
Likelihood data/ e-folding number & N =55& N=65 \\
\hline
Planck 2015 + TTTEEE (68\% CL) &- 0.50512$ < \beta < $ -0.362507 & - 0.47032$< \beta< $ - 0.30224 \\
\hline
Planck 2015 data+ TTTEEE + BAO (68\% CL)&- 0.50512$ < \beta < $ - 0.38192& - 0.47032 $< \beta< $ - 0.32512 \\
\hline
\end{tabular}
\caption{constraint on the value of $\beta$ for $N=55$ and $N=65$ using Planck likelihood}
\end{table}

                                                                          \section{inverse power-law potential}

The second potential we consider for the tachyon field is an inverse power-law potential
\begin{equation}
V(\phi)= V_{0} \phi^{-n}, \ \ \ \ \ \ \  f(\phi)= \xi_{0} \phi^{n},
\end{equation}
where $n$ is a constant. Further motivation to consider such a potential stems from Barrow's work \cite{barrow} where it is shown that the following scale factor

\begin{align}
a(t) = \exp (A t^{f}), \ \ \ \ \ \  0<f<1,  \ \ \ \ \ A>0,
\end{align}
is the solution of Friedmann equations which result in an inverse power-law potential. Such an inflationary solution evolves faster than power law inflation $(a(t) \propto t^{p}, \ p>1)$ and slower than de Sitter inflation $\left(a(t) \propto \exp( H_{dS}t),  \ H_{dS} = \mbox{const.}\right)$,  whereby known as intermediate inflation. Taking $\Gamma \propto V \propto \phi^{-n}$, we may derive the first Hubble flow function in  terms of the e-folding number
\begin{align}
&\epsilon_{1} = \frac{\frac{n}{n-4}}{N+\frac{n}{n-4}}.
\end{align}
As we can observe from the above equation $\epsilon_{1}$ is an increasing function for $n> 4$ but utilizing equations (\ref{sdf}, \ref{d2}), one finds that $f =\frac{4-n}{4}$ whereby to have $f>0$, we need $n<4$. As a result, Such a model is unphysical, similar to a pure warm tachyon inflationary model \cite{g64}. In order to make such models work,  we make the same assumption as that  considered in \cite{g64}. In fact, we suppose $\Gamma \propto \phi^{-m}$ and therefore the following Hubble and Gauss-Bonnet flow functions would result

\begin{align}
&\epsilon_{1} = \frac{\frac{n}{2m-n-4}}{N+\frac{n}{2m-n-4}}, \\&
\delta_{1} =  \frac{\beta \frac{n}{2m-n-4}}{N+\frac{n}{2m-n-4}}, \\&
\epsilon_{2} = \delta_{2} = \frac{1}{N+\frac{n}{2m-n-4}},
\end{align}
where to have $\epsilon_{1}$ as an increasing function and $f>0$, we should have $m>n+2$ similar to a pure warm tachyon inflationary model \cite{g64}. In fact the model is physically sound as long as $m>n+2$. Exploiting above equations and equations (\ref{pp6}, \ref{mm2}), one can obtain the spectral index and its running in terms of the e-folding number 

\begin{align}\label{hh1}
 n_{s}(\beta , n, m, N)& = 1- M \left(\frac{1}{N+\frac{n}{2m-n-4}}\right),
\end{align}
where
\begin{align}
M \equiv \left( \frac{1}{4}+\left(1+\frac{15}{2} \beta + \frac{4m-2}{n}\right), \left(\frac{n}{2m-n-4}\right)\right)
\end{align}
and
\begin{align}\label{hh2}
 \alpha_{s}(\beta, n, m, N)& =  - M \left(\frac{1}{N+\frac{n}{2m-n-4}}\right)^{2}.
\end{align}
Here, the spectral index varies proportional to the inverse e-folding number $N$ which means that the spectral index will be invariant for large values of the e-folding. Equations (\ref{hh1}, \ref{hh2}) are slightly more complicated than the previous relations for the spectral index and its running. In these relations, decreasing $\beta$ and $m$ and increasing $n$ will result in a substantial enhancement of the spectral index and its running. Interestingly, these quantities are also independent of dissipation coefficient $\Gamma_{0}$. The spectral index for gravitational waves is also given in  terms of the e-folding as follows

\begin{align}
n_{t}(n, m, N) = -\frac{\frac{2n}{2m-n-4}}{N+\frac{n}{2m-n-4}}.
\end{align}

\begin{figure}[H]
\includegraphics[scale=0.7]{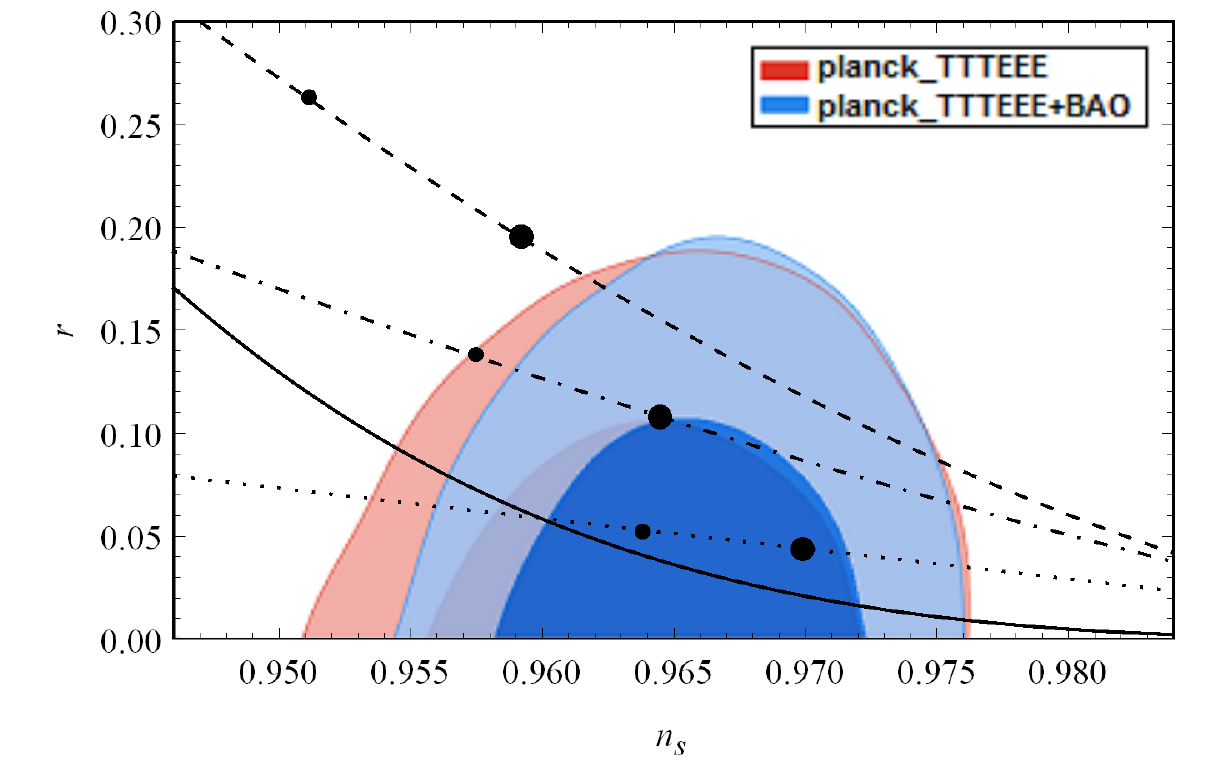} \ \ \ \ \ \ \  \includegraphics[scale=0.7]{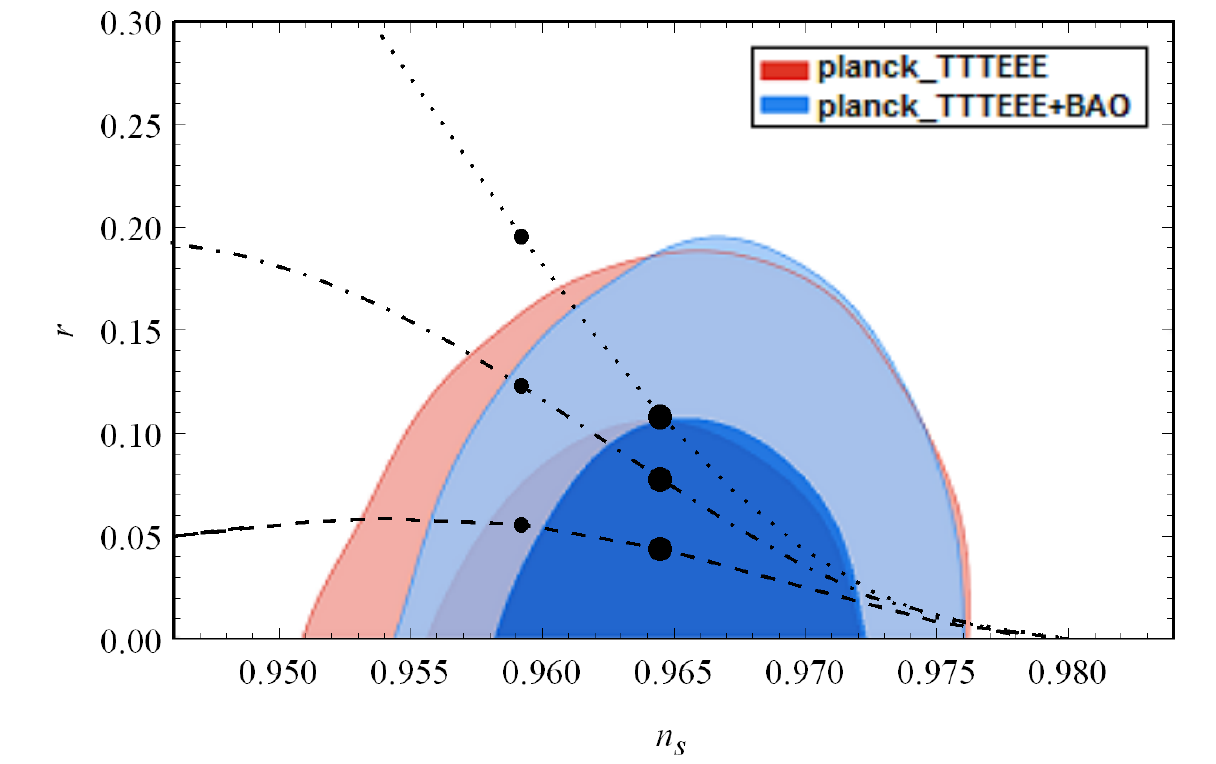}
\caption{ Two-dimensional joint marginalized constraint (68\% and 95\% confidence level) on the scalar spectral index $n_{s}$ and tensor-to-scalar ratio, $r$, including theoretical predictions where in the left panel,  dashed, dot-dashed and dotted curves denote $\beta = - 0.6$, $\beta = - 0.7$, $\beta = - 0.8$ respectively and small and large points represent $N = 50$ and $N = 60$ with $n = 3$, $m=7$, $\Gamma_{0} = 1500$. The solid curve is the prediction of warm tachyon inflation in the absence of Gauss-Bonnet coupling constant which are out of the panel for $N=50$ and $N=60$. Moreover, the number of e-folding increases moving from small to large points on each curve. In the right panel, dashed, dot-dashed and dotted curves denote $\Gamma_{0} = 500$, $ \Gamma_{0}= 1000$ and $\Gamma_{0} = 1500$ respectively and small and large points represent $\beta = - 0.6$  and $\beta = - 0.7$ with $N = 60$, $n =3$ and $m=7$. In addition, the value of $\beta$ decreases from small to large points on each curve.}
\end{figure}
We are also able to calculate tensor-to-scalar ratio in  terms of the e-folding
\begin{widetext}
\begin{align}
 r(n, m, \beta, V_{0}, \Gamma_{0}, N) = K \left(N+\frac{n}{2m-n-4}\right)^{\left(\frac{\left( -\frac{1}{4} + \frac{15}{4} \beta\right)n +\frac{3m}{2}}{-m+\frac{n}{2}+2}-\frac{3}{4}\right)}\exp \left(\frac{9}{4} \frac{(1+\beta) \left(\frac{nm}{(2m-n-4)^{2}}\right)}{\left(N+\frac{n}{2m-n-4}\right) }\right) \coth\left(\frac{k_{0}}{2 T}\right),
\end{align}
\begin{align}
K \equiv \frac{V_{0}(6 \sigma)^{\frac{1}{4}}}{96 \pi^{2}} \left(\frac{\Gamma_{0}^{2}(1+\beta)^3\left(\frac{n}{2m-n-4}\right)^{3} }{9}\right)^{\frac{1}{4}}\left(\frac{\sqrt{3}n(2m-n-4)(1+\beta)}{2 \Gamma_{0} \sqrt{V_{0}}}\right)^{\left( \frac{\left( -\frac{1}{4} + \frac{15}{4} \beta\right)n+ \frac{3m}{2}}{-m+\frac{n}{2}+2}\right)},
\end{align}
\begin{align}
T= \left(\frac{\sqrt{3 V_{0}} n^2 (1+\beta)^{2}}{4 \sigma \Gamma_{0}} \left(\frac{\sqrt{3} n (2m-n-4) (1+\beta) (N+\frac{n}{2m-n-4})}{2 \Gamma_{0} \sqrt{V_{0}}}\right)^{\left(\frac{n}{m-\frac{1}{2}n - 2}\right)}\right)^{\frac{1}{4}}.
\end{align}
\end{widetext}
One may also calculate the running of  spectral index, spectral index of gravitational waves and tensor-to-scalar ratio in terms of the spectral index as
\begin{widetext}
\begin{align}\label{tt1}
\alpha_{s}& = - M^{-1} (n_{s}-1)^{2},\\ n_{t} & = \left(\frac{2n}{2m- n-4}\right) M ^{-1} \left(n_{s}-1\right).
\end{align}
\end{widetext}

\begin{widetext}
\begin{align}\label{tt2}
 r(n, m, \beta, V_{0}, \Gamma_{0}, N)= K \left(\frac{ - M}{n_{s}-1}\right)^{\left( \frac{\left( -\frac{1}{4} + \frac{15}{4} \beta\right)n + \frac{3m}{2}}{-m+\frac{n}{2}+2} -\frac{3}{4}\right)}  \exp \left(- \frac{9}{4} {(1+\beta) \left(\frac{nm}{(2m-n-4)^{2}} \right) } M^{-1} (n_{s}-1)\right) \coth\left(\frac{k_{0}}{2 T}\right).
\end{align}
\begin{align}
T= \left(\frac{\sqrt{3 V_{0}} n^2 (1+\beta)^{2}}{4 \sigma \Gamma_{0}} \left(-\frac{\sqrt{3} n (2m- n-4) (1+\beta)  M}{2  \Gamma_{0} \sqrt{V_{0}}(n_{s}-1)}\right)^{\left(\frac{n}{m-\frac{1}{2}n - 2}\right)}\right)^{\frac{1}{4}}.
\end{align}
\end{widetext}

\begin{figure}[H]
\includegraphics[scale=0.7]{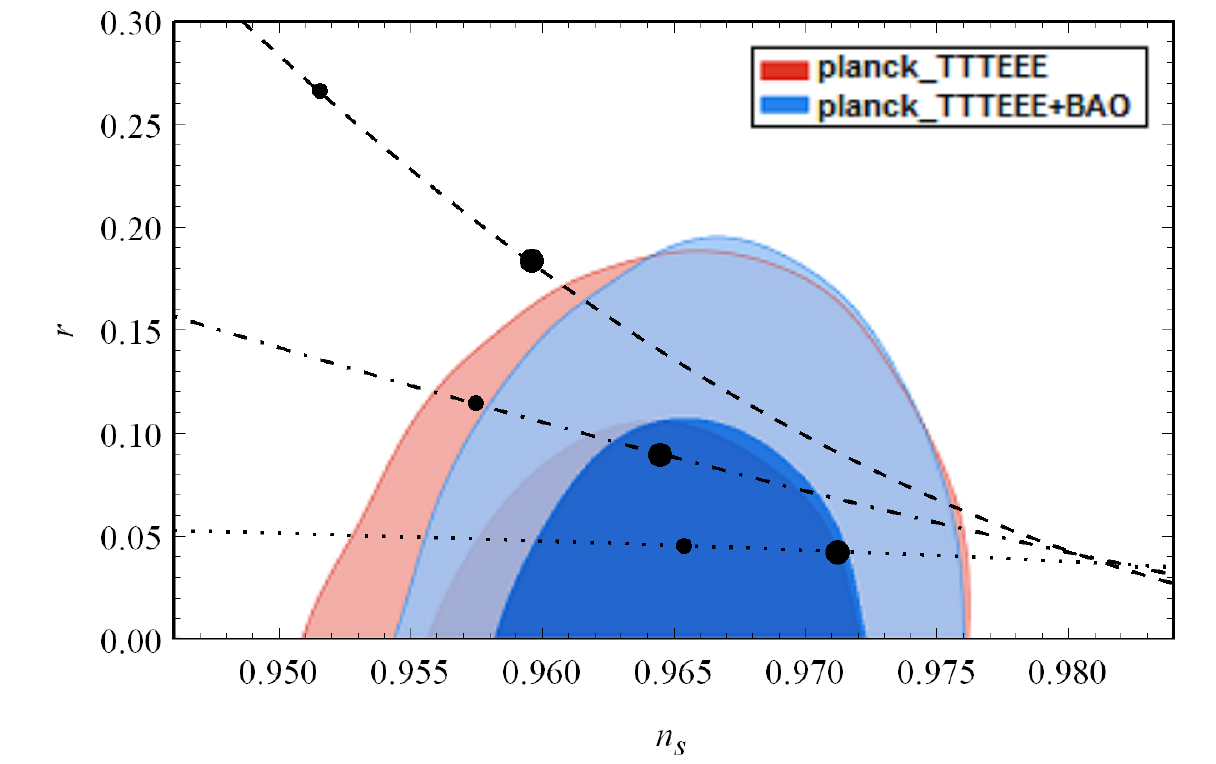} \ \  \ \ \ \ \  \includegraphics[scale=0.7]{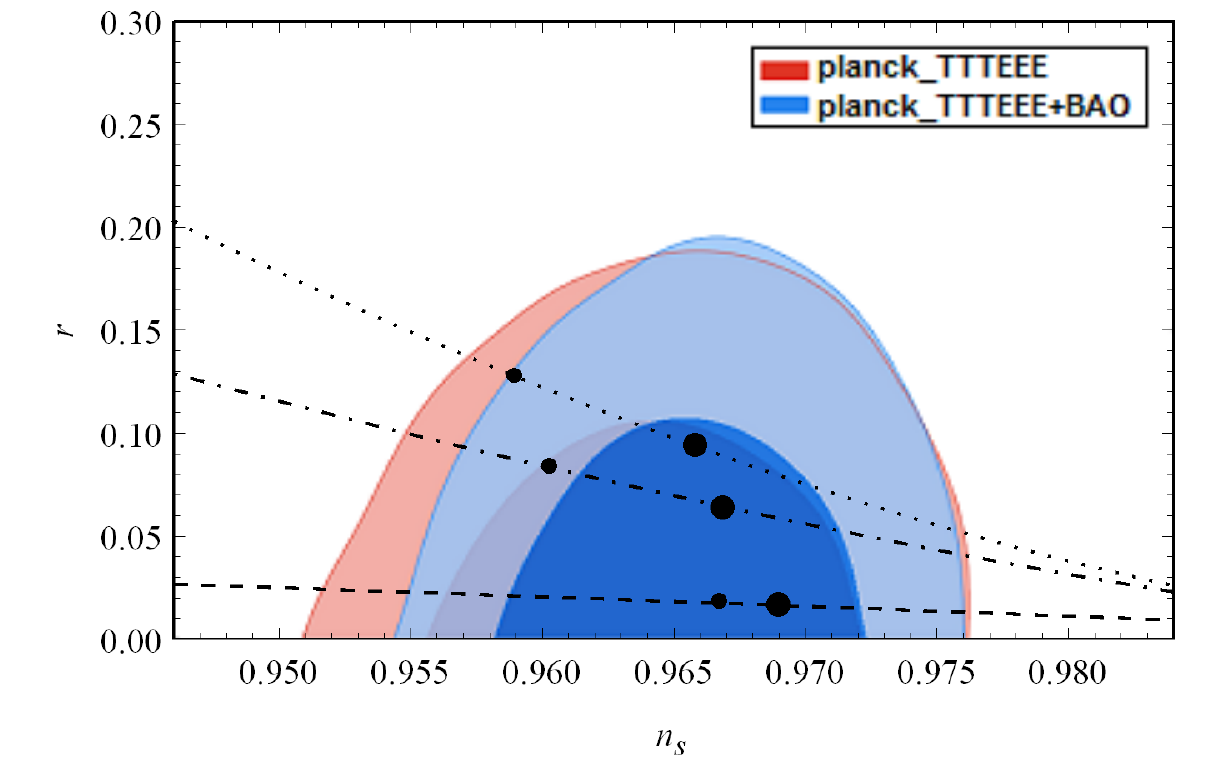}
\caption{ Two-dimensional joint marginalized constraint (68\% and 95\% confidence level) on the scalar spectral index $n_{s}$ and tensor-to-scalar ratio, $r$, including theoretical predictions where in the left panel, dashed, dot-dashed and dotted curves denote $n =2$, $n = 3$ and $n = 4$, respectively and small and large points represent $N = 50$ and $N = 60$ with $\Gamma_{0} =1200$, $\beta = - 0.7$ and $m =7$. In  the right panel, dashed, dot-dashed and dotted curves denote $m =5$, $m = 6$ and $m = 7$ respectively and small and large points represent $N = 50$ and $N = 60$ with $\Gamma_{0} =1500$, $\beta = - 0.9$ and $n =2$. Moreover, the number of e-folding increases  from small to large points on each curve in both panels.}
\end{figure}

\begin{table}[H]
\begin{center}
\begin{tabular}{ |l|l|l|l|l| }
\hline
Planck likelihood/Parameters of the model& m & n & $ \ \ \ \ \ \ \ \ \ \ \  \ \  N = 50$& $ \ \ \ \ \ \ \ \ N = 60$ \\ \hline
 & \multirow{2}{*}{ 5}& 1 & $ - 1.5451< \beta < - 0.6513 $ &$- 1.3015< \beta < - 0.2294$  \\
Planck likelihood 2015 &  & 2 &  $ - 0.90785< \beta <- 0.5233 $& $- 0.80345< \beta < - 0.3425$\\
\cline{2-5}
+ TTTEEE (95\%CL)& \multirow{2}{*}{ 6}& 2 & $ - 1.03385< \beta < - 0.53197$ & $- 0.89405 < \beta < - 0.2809$\\
&  & 3 & $ - 0.7794 < \beta < - 0.47975 $ & $- 0.6982 < \beta < - 0.33912$\\
 \hline
  & \multirow{2}{*}{ 5}& 1 & $ - 1.5451< \beta < - 0.3003 $ &$- 1.3015< \beta < - 0.1915$  \\
Planck likelihood 2015 &  & 2 & $ - 0.90785< \beta <- 0.57565 $& $- 0.80345< \beta < - 0.40525$\\
\cline{2-5}
+ TTTEEE + BAO (95\% CL)& \multirow{2}{*}{ 6}& 2 & $ - 1.03385< \beta < - 0.59165$ & $- 0.89405 < \beta < - 0.36445$\\
&  & 3 & $ - 0.7794 < \beta < - 0.52054$ & $- 0.6982 < \beta < - 0.3880$\\
 \hline

\end{tabular}
\end{center}
\caption{constraint on the value of $\beta$ for different values of $n$, $m$ and $N$ using Planck likelihoods}
\end{table}

Using relations (\ref{tt1}, \ref{tt2}) we are again able to compare theoretical predictions of this model with the two-dimensional joint marginalized constraint. In the left panel of figure 2, we show three different values of $\beta$ and variation of the e-folding number where for fixed values of e-folding, decreasing the value of $\beta$  horizontally shifts the spectral index and vertically shifts tensor-to-scalar ratio $r$. In fact, decreasing the value of $\beta$ results in a substantial enhancement in the value of $n_{s}$ and $r$. In addition, setting $\xi_{0}$ to zero,  it is seen from observational constraint that this is not in agreement with observational data for fixed values of parameters of the model in figure 2. The right panel of figure 2 shows three different values of dissipation coefficient amplitude and variation of $\beta$ where for fixed values of $\beta$, decreasing the value of dissipation coefficient amplitude  shifts  tensor-to-scalar ratio vertically but keeps the spectral index invariant. Subsequently, decreasing the value of $\Gamma_{0}$ results in an enhancement in the value of $r$ and does not change the value of the spectral index.  The left panel of figure 3 illustrates three different values of $n$ and variation of e-folding number where for a fixed value of the e-folding, increasing the value of $n$ horizontally shifts the spectral index and  vertically shifts tensor-to-scalar ratio. In fact, increasing the value of $n$ results in an observable enhancement in the value of $n_{s}$ and $r$. On the other hand, the right panel of figure 3 shows three different values of $m$ and variation of the e-folding number where for a fixed value of the e-folding, decreasing the value of $m$ horizontally shifts the spectral index and vertically shifts tensor-to-scalar ratio. In fact, decreasing the value of $m$ results in a substantial enhancement in the value of $n_{s}$ and $r$.

Finally, Using equations (\ref{hh1}, \ref{hh2}) one can find some constraints on $\beta$ for different values of $n$ and $m$. These results are summarized in table II for $N = 50$ and $N= 60$.

                                             \section{Exponential potential with power-law Gauss-Bonnet coupling}

 Let us now consider a further general form for the potential and Gauss-Bonnet function
\begin{equation}
V(\phi) = V_{0} e^{-\alpha \phi}, \ \ \  \ \ \ f(\phi) = \xi_{0} \phi^{n}.
\end{equation}
As one cannot  integrate equation (\ref{pp5}) explicitly the results will be obtained numerically. Again, if we consider $\Gamma \propto V$, we find  the first flow function as
\begin{align}
\epsilon_{1}&= \sqrt{\frac{3}{2V_{0} \Gamma_{0}^{2}}} { ( \alpha^{2} e^{\alpha \phi} + n \alpha \beta \phi^{n-1}) e^{- \frac{1}{2}\alpha \phi}}, \\ \delta_{1}& =  \sqrt{\frac{3}{4V_{0} \Gamma_{0}^{2}}}\left(\beta n \alpha  e^{\alpha \phi} + \beta^{2} n^{2} \phi^{n-1}\right) \phi^{n-1} e^{- \frac{3}{2} \alpha \phi}.
\end{align}
Using the above equations and definition for flow functions we can easily derive the second flow functions as a function of $\phi$. Inflation will then end when $\epsilon_{1} \simeq 1$ and solving this equation enables us to find the value of $\phi$ at the end of inflation. By setting the e-folding to 50, 60 or 70 we can numerically integrate and obtain the value of $\phi$ at the Hubble crossing time. Using this value and equations (\ref{pp6}, \ref{mm2}, \ref{mm3}, \ref{mm4}), we can find the value of the spectral index, its running and tensor-to- scalar ratio for different values of free parameters. We have also  plotted  tensor-to-scalar ratio, its running and the spectral index for gravitational waves versus scalar spectral index.

In the left panel of figure 4, we show three different values of dissipation coefficient amplitude and  variation of $\xi_{0}$ where for fixed values of $\xi_{0}$, decreasing the value of $\Gamma_{0}$  horizontally shifts the spectral index and  shifts tensor-to-scalar ratio $r$ vertically. Therefore, decreasing the value of $\Gamma_{0}$ results in a great enhancement in the values of $n_{s}$ and $r$. The right panel in figure 4 is plotted for three different values of $\xi_{0}$ and variation of $\alpha$  for fixed values of $\xi_{0}$. We see that decreasing the value of $\alpha$ horizontally shifts the spectral index and  shifts tensor-to-scalar ratio $r$ vertically. In fact, decreasing the value of $\alpha$ results in an observable enhancement in the value of $n_{s}$ and $r$. Throughout our calculations we have taken $\sigma = 1$.

\begin{table}[H]
\begin{center}
\begin{tabular}{ |l|l|l|l| }
\hline
 \ \ \ n \ \ \ & the number of e-folding (N) & $\ \ \ \ \ \ \ \ \ \ \ \ \  \xi_{0}$ \\ \hline
\multirow{2}{*}{\ \ \ 2 \ \ \ }&  \ \ \ \ \ \ \ \ \ \ \ \ \ \  55 & $ - 1.690< \xi_{0} < - 0.816$ \\
 &  \ \ \ \ \ \ \ \ \ \ \ \ \ \   65 & $ -1.791 < \xi_{0} < -0.577 $ \\
 \hline
 \multirow{2}{*}{\ \ \ 3 \ \ \ }&  \ \ \ \ \ \ \ \ \ \ \ \ \ \ 55 & $ - 0.450< \xi_{0} < - 0.295$ \\
 & \ \ \ \ \ \ \ \ \ \ \ \ \ \ 65 & $- 0.453 < \xi_{0} < - 0.317$ \\
 \hline
 \multirow{2}{*}{\ \ \ 4 \ \ \ }&  \ \ \ \ \ \ \ \ \ \ \ \ \ \ 55 & $ - 0.097< \beta < - 0.087$ \\
 & \ \ \ \ \ \ \ \ \ \ \ \ \ \  65 & $- 0.097 < \xi_{0} < - 0.085$ \\
 \hline
\end{tabular}
\caption{The range of $\xi_{0}$ for $\Gamma_{0} =180$, $\alpha= 0.9$, $V_{0}=0.6$ and different values of $n$ using Planck 2015+ TTTEEE+ BAO likelihood (95 \%CL).}
\end{center}
\end{table}

In so far as mentioned before, the consistency relation is violated in the context of warm inflation and is modified. This then gives us the opportunity to utilize four cosmological quantities, namely $n_{s}, \alpha_{s}, r$ and $n_{t}$ as constraint equations. These equations would then enable us to numerically fix three parameter of the model and find constraints on the remaining one. Therefore, our constraint on the values of $\xi_{0}$ is summarized in table III for Planck 2015+ TTTEEE+ BAO likelihood data. In fact, such ranges for $\xi_{0}$ simultaneously satisfy all the constraints on $n_{s}, n_{t}, \alpha_{s}$ and $r$.

\begin{figure}[H]
\includegraphics[scale=0.7]{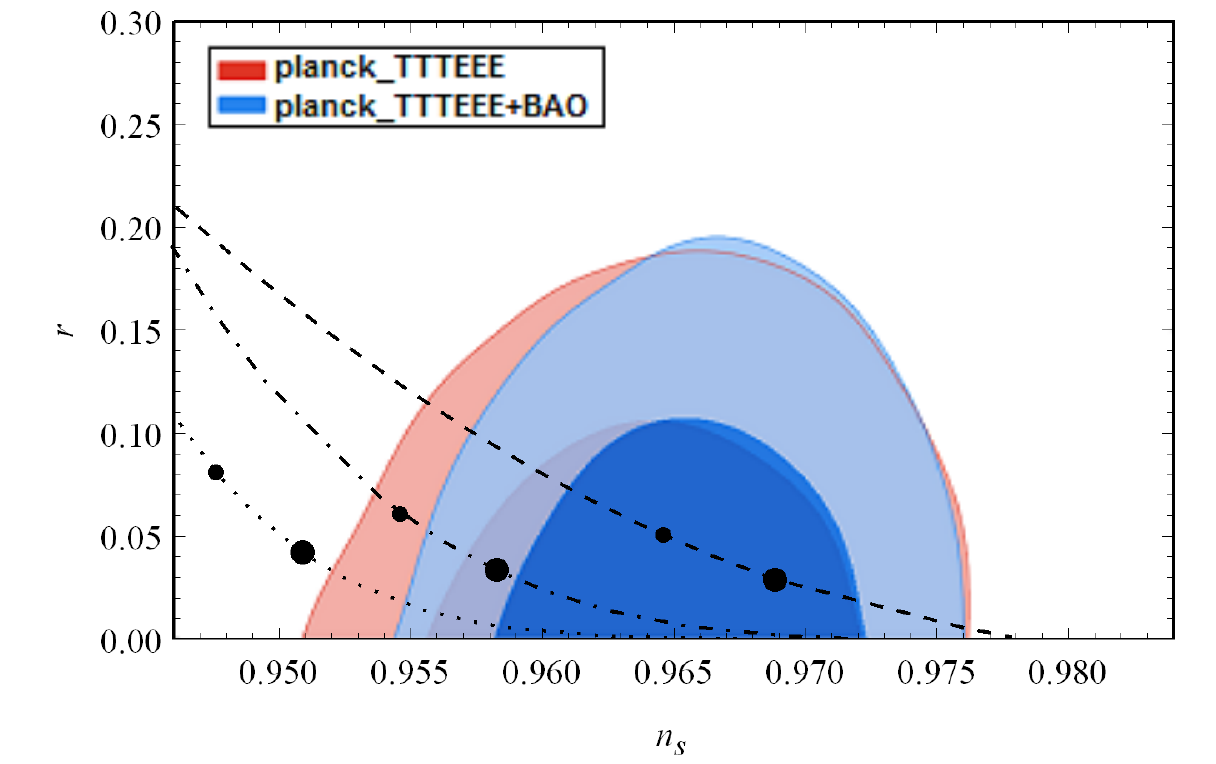}  \ \ \ \ \ \ \  \includegraphics[scale=0.7]{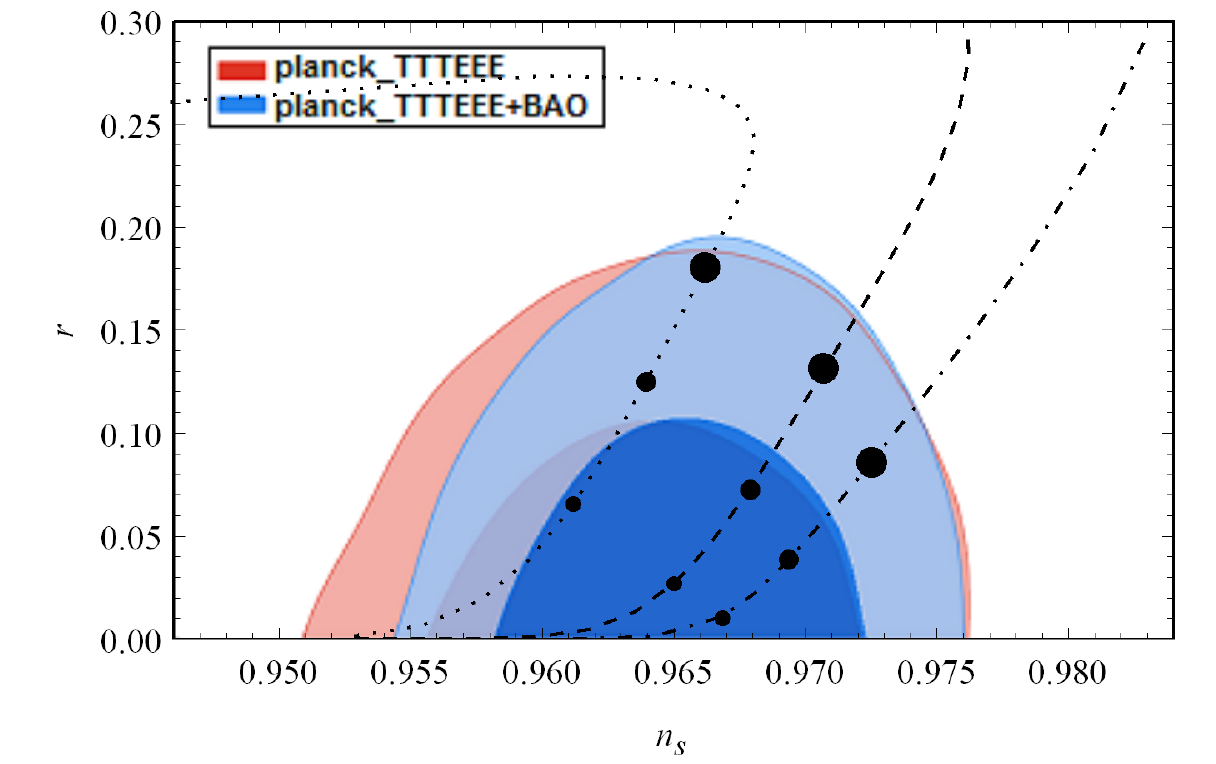}
\caption{ Two-dimensional joint marginalized constraint (68\% and 95\% confidence level) on the scalar spectral index $n_{s}$ and tensor-to-scalar ratio, $r$, including theoretical predictions where in the left panel, dashed, dot-dashed and dotted curves denote $\Gamma_{0} = 100$, $\Gamma_{0} = 150$ and $\Gamma_{0} =200$ respectively and  small and large points represent $ \xi_{0}  = - 0.3$ and $\xi_{0} = - 0.35$ with $n = \frac{3}{2}$, $\alpha  =0.9$ and $V_{0} = 1$. In addition, the value of $\xi_{0}$ decreases  from small to large points on each curve and the case $\xi_{0} = 0$ is out of panel for three values of $\Gamma_{0}$. In the right panel, dotted, dashed and dot-dashed curves denote $\xi_{0} = - 0. 35$, $\xi_{0} = - 0.45$ and $\xi_{0} = -0.55$ respectively and small, medium and large points represent $\alpha = 0.6$, $\alpha = 0.7$ and $\alpha = 0.8$ with $V_{0} = 0.8$, $n = 2$ and $\Gamma_{0} = 100$. Moreover, the value of $\alpha$ increases moving from small to large points on each curve.}
\end{figure}

                                                                              \section{summary and conclusions}

In recent years, warm inflationary scenarios have attracted great attention as  complementary versions of  conventional inflation \cite{q9}. The reason is that these scenarios inherit the properties of standard inflation and are also able to avoid the reheating period, solving the so-called eta problem and alleviate the initial condition problem. Such appealing characteristics were our motivation to study tachyon inflation in the context of a warm inflationary scenario modified by adding a low-energy stringy correction.

The general form of the modified spectral index and power spectrum were derived in terms of generalized slow-roll parameters in a high dissipation regime. In the absence of a Gauss-Bonnet coupling constant ($\xi_{0}=0$) the model is theoretically consistent with warm-tachyon inflation and  for $\xi_{0}=0$ and $\Gamma=0$ the cosmological perturbations of the model coincide with that of the cold inflation. We started by analytically solving our model for two potentials, ($V(\phi)= V_{0} e^{-\alpha \phi}$) and ($V(\phi) = V_{0} \phi^{-n}$), which satisfy the properties of a tachyon field and  lead to theoretically convincing results in high dissipation regimes. We were also able to find some ranges for $\beta$ for which our model is consistent with the recent data, summarized in TABLE I and TABLE II. Next, we further considered  general functions and numerically solved our model in order to find constraint on the parameters of the model. Since tensor-to-scalar ratio gets modified in the context of warm inflation it gives us the opportunity to utilize four parameters at our disposal, namely $n_{s}, n_{t}, r, \alpha_{s}$ as four constraint equations in order to reduce the number of parameters of the model and found some ranges for $\xi_{0}$ for which the model is  consistent with a $95\% $ confidence level. These results have been summarized in TABLE III. In fact, the presence of a Gauss-Bonnet term adds one degree of freedom to our system but violation of consistency relation allows us to independently utilize the aforementioned space parameters as constraint equations. Therefore, the Gauss-Bonnet term gives our model further freedom to be fixed by observation, although, recently released Planck data put tight bounds on the tensor-to-scalar ratio($r<0.12$). In general  we found that decreasing the value of the Gauss-Bonnet coupling  enhances the value of the spectral index and tensor-to-scalar ratio which causes the model to become inconsistent with observation for positive values of $\xi_{0}$. In fact, the Gauss-Bonnet coupling  controls  termination of inflation and is in agreement with observation even for steeper potentials. Furthermore, the model has the potential to cover an spectrum  running from blue ($n_{s}>1$) to red ($n_{s}<1$) for some ranges of $\xi_{0}$. Indeed, there is a further freedom on the range of the spectral index. This property usually arises in models where the inflaton field undergoes interaction with other fields or a dissipative factor is present. In particular, we anticipate that the future data would accord us  a more  accurate understanding of $\alpha_{s}$ and the power spectrum.

As a final remark, since the model described above presents a change in the source of initial cosmological fluctuations, it may have a substantial effect on baryogenesis process, graviton production,  evolution of matter in the intermediate epoch which  deserves investigation.  In this paper, we have not addressed non-Gaussianity of cosmological perturbations but hope to present such an analysis in a separate work.

                                                                                  \section{acknowledgement}

We would like to acknowledge the use of  cosmoMC exploring engine and thank Antony Lewis and his colleagues for providing a helpful forum which was instrumental in using this package.


\end{document}